\documentstyle[aps,prb,epsfig,floats]{revtex}

\begin{document}
\draft


\twocolumn[\hsize\textwidth\columnwidth\hsize\csname@twocolumnfalse%
\endcsname
\title{Theory of anomalous magnon softening in ferromagnetic manganites}
\author{G.\ Khaliullin \protect\cite{ADD}}
\address{Max-Planck-Institut f\"ur Festk\"orperforschung,
Heisenbergstr.\ 1, D-70569 Stuttgart, Germany}
\author{R.\ Kilian}
\address{Max-Planck-Institut f\"ur Physik komplexer Systeme,
N\"othnitzer Strasse 38, D-01187 Dresden, Germany}
\date{\today}
\maketitle


\begin{abstract}
In metallic manganites with low Curie temperatures,
a peculiar softening of the magnon spectrum close to the
magnetic zone boundary has experimentally been observed. 
Here we present a theory of the renormalization of the
magnetic excitation spectrum in colossal magnetoresistance 
compounds. The theory is based 
on the modulation of magnetic exchange bonds by the orbital 
degree of freedom of double-degenerate $e_g$ electrons. The 
model considered is an orbitally degenerate double-exchange 
system coupled to Jahn-Teller active phonons which we treat
in the limit of strong onsite repulsions. Charge and coupled 
orbital-lattice fluctuations are identified as the main 
origin of the unusual softening of the magnetic spectrum.
\end{abstract}

\pacs{PACS number(s): 75.30.Ds, 75.30.Et}
\rule{0mm}{8mm}]


\section{Introduction}

The motion of charge carriers in the metallic phase of manganites
establishes a ferromagnetic interaction between spins on neighboring sites.
According to the conventional theory of double 
exchange,\cite{ZEN51,AND55,DEG60,KUB72,FUR96} the spin dynamics of the 
ferromagnetic state that evolves at temperatures below the Curie 
temperature $T_C$ is expected to be of nearest-neighbor Heisenberg type. 
This picture seems to be indeed reasonably accurate for manganese oxides
with large values of $T_C$, i.e., for compounds whose 
ferromagnetic metallic phase sustains up to rather high 
temperatures.\cite{PER96}
However, recent experimental studies indicate marked deviations
from this canonical behavior in compounds with low values
of $T_C$. Quite prominent in this respect are measurements of the spin 
dynamics of the ferromagnetic manganese oxide 
Pr$_{0.63}$Sr$_{0.37}$MnO$_3$:\cite{HWA98} While exhibiting 
conventional Heisenberg behavior at small 
momenta, the dispersion of magnetic excitations (magnons) shows curious 
softening at the boundary of the Brillouin zone. 
This observation is of high importance as it indicates that some specific 
feature of magnetism in manganites has yet to be identified.

A comparison of the magnetic behavior of different manganese oxides 
further highlights the shortcomings of the double-exchange theory:
Assuming the magnon dispersion to be of Heisenberg type,
a small-momentum fit to a quadratic dispersion relation 
$\omega_{\bbox{q}} = D q^2$ yields the spin-wave stiffness $D$; 
in a conventional 
Heisenberg system the spin-wave stiffness scales with the strength
of magnetic exchange bonds $D \propto J$. Since the latter
also controls the Curie temperature $T_C \propto J$, the ratio of 
$D$ and $T_C$ is expected to be a universal constant.
Manganites, on the other hand, exhibit a pronounced 
deviation from this behavior: $D/T_C$ increases significantly as one 
goes from compounds with high to compounds with low values of 
$T_C$.\cite{FER98} The presence of an additional mechanism that 
controls the magnetic behavior of manganites is to be inferred.

\begin{figure}
\centering
\setlength{\unitlength}{0.22\linewidth}
\rule{0.15\unitlength}{0\unitlength}
\begin{picture}(1,1.92)
\put(0,0){\epsfig{file=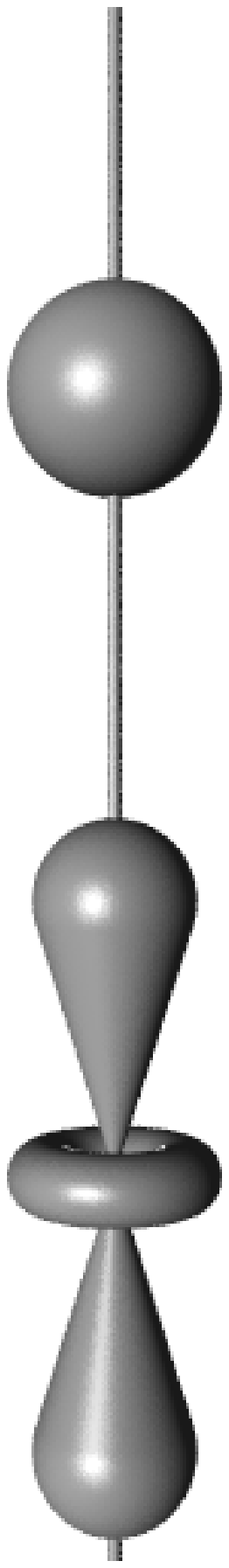,width=\unitlength,clip=}}
\put(-0.15,0.6){\epsfig{file=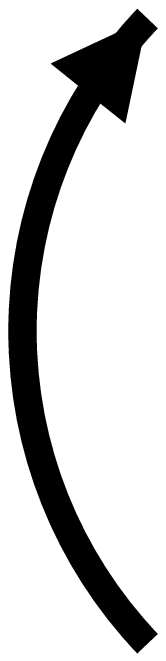,width=0.4\unitlength}}
\put(0.1,-0.35){\fbox{$J_{\text{DE}}^c \propto t$}}
\end{picture}
\rule{0.4\unitlength}{0\unitlength}
\rule{0.45\unitlength}{0\unitlength}
\begin{picture}(1,2.14)
\put(0,0){\epsfig{file=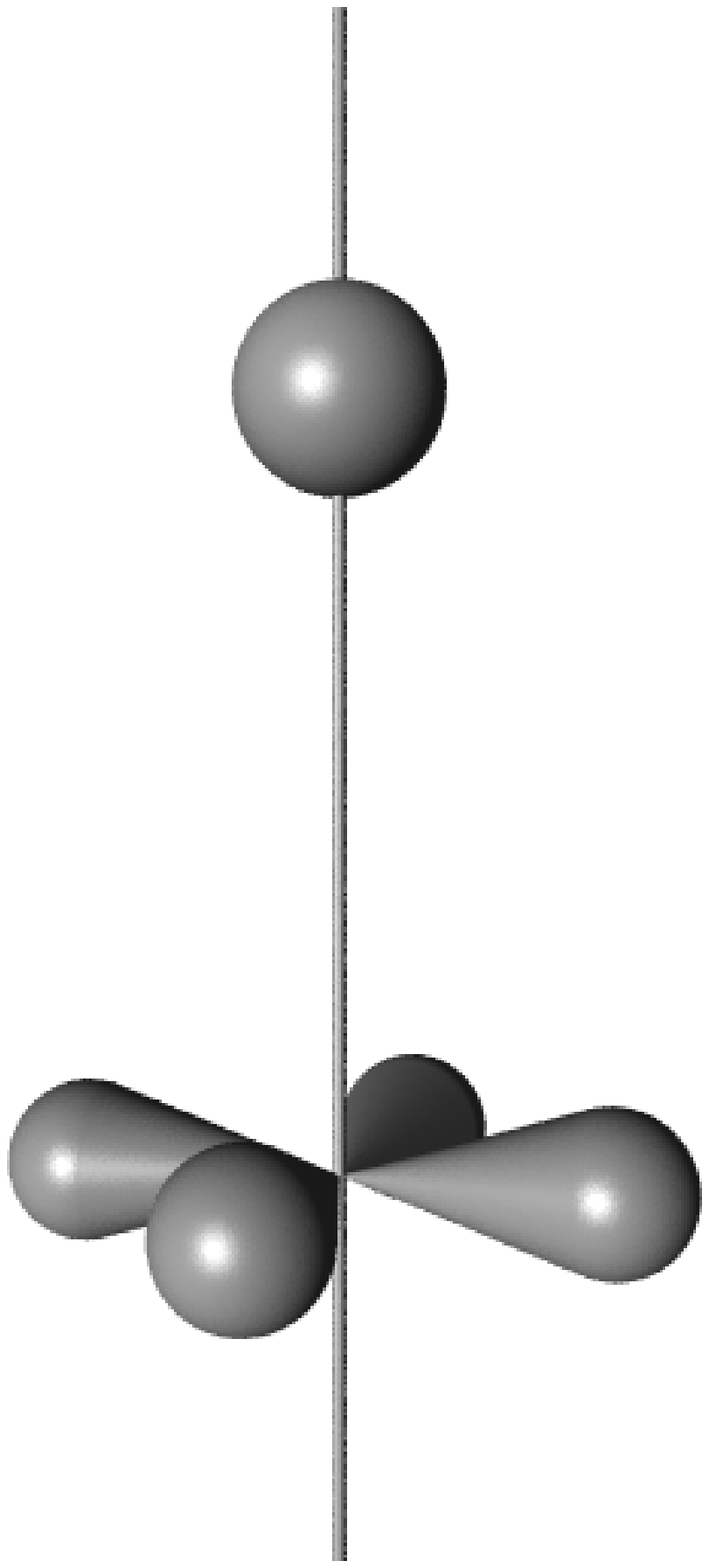,width=\unitlength,clip=}}
\put(-0.45,0.6){\epsfig{file=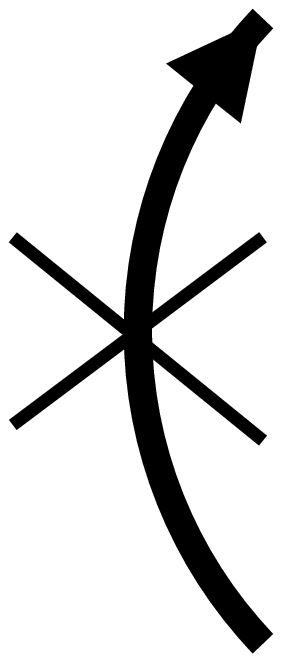,width=0.4\unitlength}}
\put(0.1,-0.35){\fbox{$J_{\text{DE}}^c = 0$}}
\end{picture}
\rule{0.15\unitlength}{0\unitlength}\\[0.65\unitlength]
\caption{The $e_g$-electron transfer amplitude, which controls the
double-exchange interaction $J_{\text{DE}}$, strongly depends on
the orbital orientation: Along the $z$ direction, e.g., $d_{3z^2-r^2}$ 
electrons (left) can hop into empty sites denoted by a sphere, 
while the transfer of $d_{x^2-y^2}$ electrons (right) is forbidden.}
\label{FIG:MBD}
\end{figure}

\begin{figure}
\centering
\epsfig{file=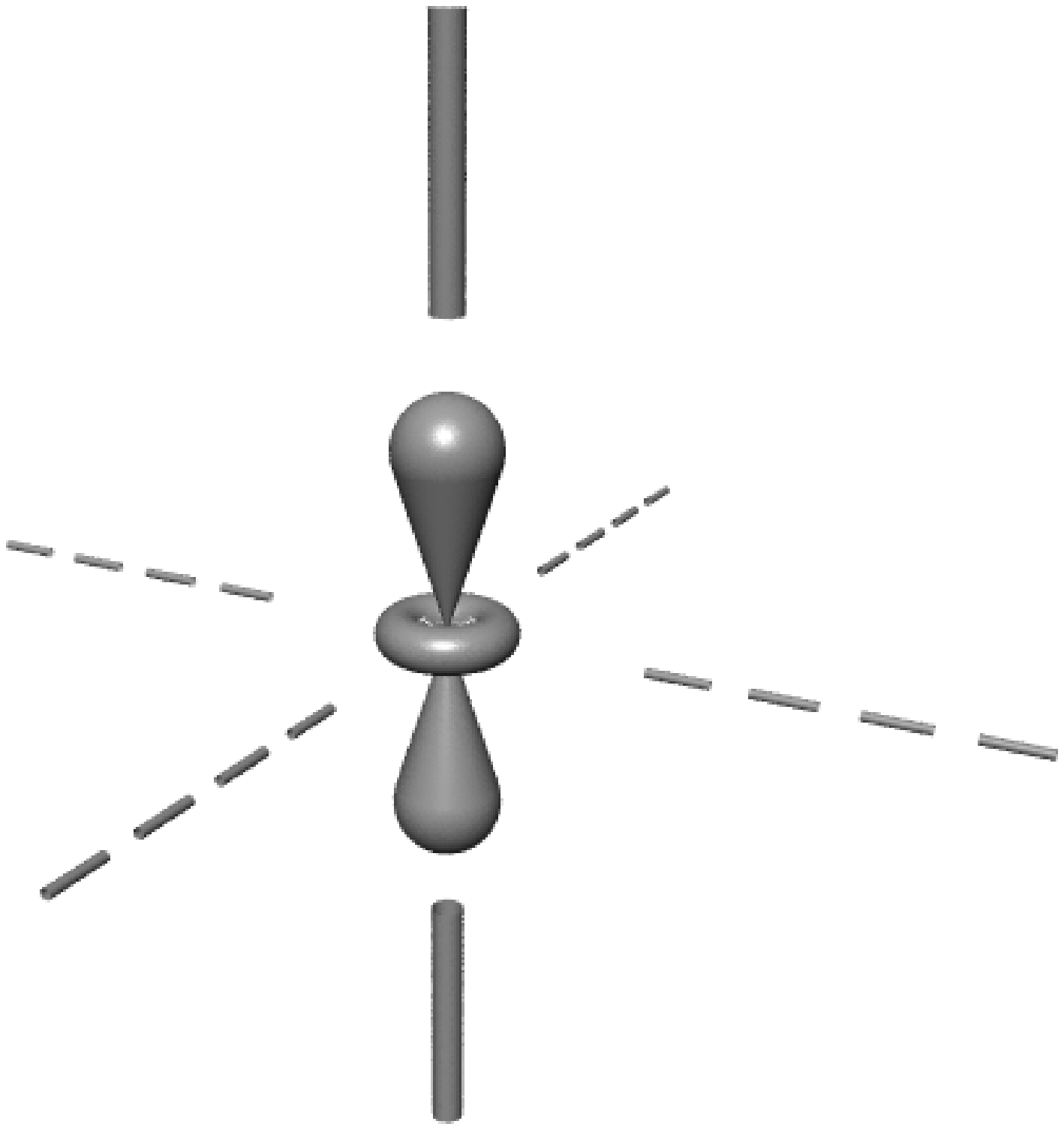,width=0.49\linewidth,clip=}
\hfill
\epsfig{file=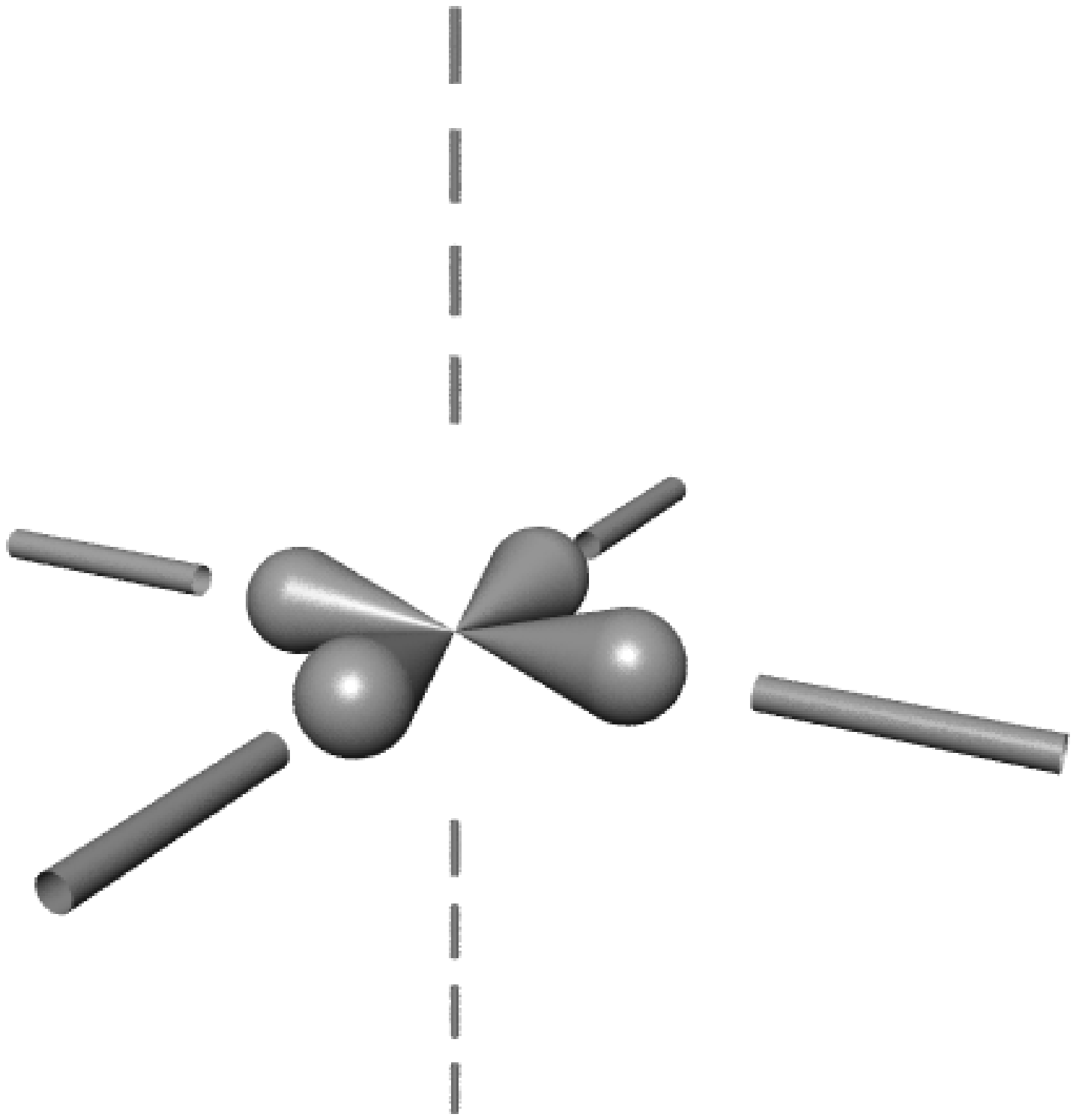,width=0.49\linewidth,clip=}\\[6pt]
\caption{Fluctuation of magnetic exchange bonds: Full lines
denote active bonds, dashed lines inactive ones.}
\label{FIG:EXB}
\end{figure}

In the present paper, we propose a mechanism to explain the
above peculiar magnetic properties of ferromagnetic manganites.
Our basic idea is the following: The strength of the ferromagnetic 
interaction at a given bond strongly depends on the orbital
quantum number of $e_g$ electrons (see Fig.\ \ref{FIG:MBD})~---
along the $z$ direction, e.g., only electrons in $d_{3z^2-r^2}$ orbitals can 
hop between sites and hence can participate in double-exchange processes; 
the transfer of $d_{x^2-y^2}$ electrons is blocked due to the vanishing overlap
with O $2p$ orbitals located in-between two 
neighboring Mn sites. Temporal fluctuations of $e_g$ orbitals may thus
modulate the magnetic exchange bonds (see Fig.\ \ref{FIG:EXB}), thereby
renormalizing the magnon dispersion. Short-wavelength magnons are
most sensitive to these local fluctuations and are affected most strongly. 
Quantitatively the modulation of exchange bonds is controlled by the 
characteristic time scale of orbital fluctuations: If the typical 
frequency of orbital fluctuations is higher than the one of spins 
fluctuations, the magnon spectrum remains mostly unrenormalized~---
the orbital state then effectively enters the spin dynamics only on time 
average which restores the cubic symmetry of exchange bonds.
On the other hand, if orbitals fluctuate slower than spins, the 
renormalization of the magnon spectrum is most pronounced~--- the 
anisotropy imposed upon 
the magnetic exchange bonds by the orbital degree of freedom 
now comes into play. The presence of Jahn-Teller 
phonons enhances this effect by quenching the dynamics of orbitals.
The suppression of fluctuations becomes almost complete
as orbitals begin to order, resulting in a distinct softening of 
magnons which we interprete as a precursor effect of static orbital 
order.
 
In the following, we calculated the dispersion of one-magnon excitations
at zero temperature. We start from an orbitally degenerate Hubbard model
that comprises the strongly correlated nature of the Mn $3d$ electrons
and the physics of double exchange. The metallic motion 
of charge carriers establishes magnetic double-exchange
bonds which are found to be further contributed to 
by virtual superexchange processes. Both types of exchange
interaction are of ferromagnetic nature in the orbitally degenerate 
system subject to a strong Hund's coupling.
Employing a $1/S$ expansion of spin and an orbital-liquid scheme
\cite{ISH97,KIL98} to handle correlation effects, three 
different mechanisms are analyzed with respect to their capability to
renormalize the magnon spectrum: scattering of magnons on orbital
fluctuations, on charge fluctuations, and on phonons. Within this
picture we can successfully reproduce the experimentally observed 
softening of the magnon dispersion. Furthermore we predict the 
renormalization effect to become dramatic as static order in the 
orbital-lattice sector is approached. We note that the 
renormalization of the magnetic excitation spectrum by optical 
phonons has recently been investigated by Furukawa.\cite{FUR99}


\section{Magnetic Exchange Bonds}
\label{SEC:EXB}

The main aspects of the physics of manganites, i.e., the
correlated motion of itinerant $e_g$ electrons and the ferromagnetic
interaction of $e_g$ spins with a background of localized core spins, is 
captured by the following orbitally degenerate Hubbard model:
\begin{eqnarray}
H_{\text{Hub}} &=& -\sum_{\langle ij \rangle_{\gamma}}\sum_{s\alpha\beta} 
t_{\gamma}^{\alpha\beta}\left(c^{\dagger}_{is\alpha}c_{js\beta}
+\text{H.c.}\right)- J_H \sum_i \bbox{S}^c_i \bbox{s}_i\nonumber\\*
&& + \sum_i\sum_{\alpha}Un_{i\uparrow\alpha}n_{i\downarrow\alpha}
\nonumber\\*
&& + \sum_i{\sum_{\alpha\ne\beta}}'\left(U'-J_H\hat{P}\right)
n_{i\alpha}n_{i\beta},
\label{HBB}
\end{eqnarray}
with $\hat{P}=(\bbox{s}_{i\alpha} \bbox{s}_{i\beta}+\frac{3}{4})$.
The first term in Eq.\ (\ref{HBB}) describes the intersite transfer
of electrons within degenerate $e_g$ levels. Here, $c^{\dagger}_{is\alpha}$
creates an $e_g$ electrons with spin and orbital quantum numbers $s$ and 
$\alpha$/$\beta$, respectively.
The spatial direction of bonds is specified by $\gamma\in\{x,y,z\}$.
One of the important features of the orbitally degenerate model is the 
nondiagonal structure of the transfer matrices\cite{KUG82}
\[
t_{x/y}^{\alpha\beta} = t\left(
\begin{array}{cc} 
1/4 & \mp\sqrt{3}/4 \\ 
\mp\sqrt{3}/4 & 3/4
\end{array}
\right),\quad
t_z^{\alpha\beta} = t \left(
\begin{array}{cc} 
1 & 0 \\ 
0 & 0
\end{array}
\right),
\]
where we have chosen a representation with respect to the orbital basis
$\{|3z^2-r^2\rangle,|x^2-y^2\rangle\}$. The second term of 
Eq.\ (\ref{HBB}) describes the Hund's coupling between the 
itinerant $e_g$ electrons and the localized core spins $\bbox{S}^c_i$;
the magnitude of this coupling is $J_H$. The spin operator 
$\bbox{s}_{i\alpha}$ acts on $e_g$ electrons in orbitals $\alpha$, 
while $\bbox{s}_i=\sum_{\alpha}\bbox{s}_{i\alpha}$ denotes
the total $e_g$ spin at a given site.
Finally, the last two terms in model (\ref{HBB}) account for the 
intra- (inter-) orbital Coulomb interaction $U$ ($U'$) and the Hund's 
coupling between $e_g$ electrons in doubly occupied states.
$n_{is\alpha}$ is the number operators of $e_g$ 
electrons in the state defined by $s$ and $\alpha$, and
$n_{i\alpha} = \sum_s n_{i\alpha}$. Double counting is
excluded from the primed sum in the last term of Eq.\ (\ref{HBB}).

In analogy to the transformation from a conventional Hubbard to $t$-$J$ 
model, Eq.\ (\ref{HBB}) can be projected onto the part of the Hilbert space 
with no double occupancies in the limit of strong onsite repulsions 
$U \gg t$ and $(U'-J_H)\gg t$. Doubly occupied states are then allowed
only in virtual superexchange processes. Due to the presence of Hund's 
coupling, the energy level of these virtual states depends on the spin 
orientation of
core and $e_g$ spins~--- a rich multiplet structure follows.\cite{FEI99}
The problem considerably simplifies in the limit of large Hund's 
coupling $U,U' \gg J_H \gg t$ which we believe to be realistic to 
manganites: Transitions to the lowest-lying intermediate state with 
energy $U_1=U'-J_H$ in which core and $e_g$ spins are in a high-spin 
configuration then dominate; doubly occupied sites with different 
spin structures lie higher by an energy of the order of $\propto J_H$ 
and can be neglected. We hence obtain the following $t$-$J$ Hamiltonian:
\begin{eqnarray}
H_{tJ} &=& -\sum_{\langle ij \rangle_{\gamma}}\sum_{s\alpha\beta} 
t_{\gamma}^{\alpha\beta}\left(\hat{c}^{\dagger}_{is\alpha}\hat{c}_{js\beta}+
\text{H.c.}\right)- J_H\sum_i \bbox{S}_i^c \bbox{s}_i\nonumber\\* 
&& - J_{\text{SE}} \sum_{\langle ij \rangle_{\gamma}}
\left(\case{1}{4}-\tau_i^{\gamma}\tau_j^{\gamma}\right)
\left[\bbox{S}_i\bbox{S}_j + S(S+1)\right]n_i n_j.\nonumber\\*
\label{HTJ}
\end{eqnarray}
The first two terms in Eq.\ (\ref{HTJ}) describe the 
double-exchange mechanism in the limit of strong onsite repulsions.
All double occupancies of $e_g$ electrons are projected out by 
the constrained operators $\hat{c}^{\dagger}_{is\alpha} = 
c^{\dagger}_{is\alpha} (1-n_i)$ which act only on empty sites. 
The third term in Eq.\ (\ref{HTJ}) describes the superexchange 
interaction between singly occupied sites. The 
strength of this interaction
is controlled by $J_{\text{SE}} = (2t^2/U_1)[S(2S+1)]^{-1}$, 
where $S$ denotes the total onsite spin of $3d$ electrons. It is 
important to note that in the present model with large $J_H$, 
superexchange is of {\it ferromagnetic} nature. This stems from 
the fact that Hund's coupling forbids any double occupancy of a single
$e_g$ orbital. Pauli's exclusion principle, which is responsible for the 
antiferromagnetic nature of conventional superexchange, is therefore 
ineffective in dictating the spin structure of the virtual state. 
Rather, the spin orientation in the intermediate state is controlled 
by Hund's coupling which favors a ferromagnetic alignment of spins.
The superexchange term in Eq.\ (\ref{HTJ}) exhibits yet another
peculiar feature: The amplitude of superexchange processes
depends on the orbital states of the $e_g$ electrons involved.
This information enters via the orbital pseudospin operators
\[
\tau_i^{x/y} = -\case{1}{4}\left(\sigma_i^z\pm\sqrt{3}\sigma_i^x\right),
\quad \tau_i^z = \case{1}{2}\sigma_i^z,
\]
where the Pauli matrices $\sigma_i^{x/z}$ act on the orbital subspace;
the factor $(\frac{1}{4}-\tau_i^{\gamma} \tau_j^{\gamma})$ in 
Eq.\ (\ref{HTJ}) accounts
for the specific nondiagonal structure of the transfer matrices 
$t_{\gamma}^{\alpha\beta}$ and ensures that no double occupancy 
of a single orbital occurs which would be forbidden by Pauli's
exclusion principle and the large Hund's coupling.
We finally note that superexchange processes in an orbitally
degenerate system have also been studied
by Feiner and Ole\'s.\cite{FEI99} In the limit of large
$J_H$ the expression obtained by these authors maps onto
the superexchange term of Eq.\ (\ref{HTJ}) for the special
case $S=2$.

In the following, double-exchange and superexchange interactions 
which are jointly responsible for ferromagnetism in manganites are 
discussed in more detail.


\subsection{Double-Exchange Bonds}

\begin{figure}
\centering
\setlength{\unitlength}{0.6\linewidth}
\begin{picture}(1,0.8)
\put(0,0){\epsfig{file=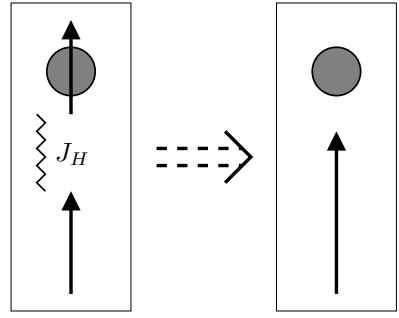,width=\unitlength}}
\put(0.12,0.39){$J_H$}
\end{picture}\\[6pt]
\caption{The itinerant $e_g$ spin (top left) interacts with 
the localized core spins (bottom left) via Hund's coupling. In the limit 
$J_H \gg t$, the former can be separated from the orbital 
and charge degrees of freedom of the $e_g$ electron (circle) and 
can be absorbed into the total spin (bottom right).}
\label{FIG:ABS}
\end{figure}

We begin by analyzing the kinetic term of Hamiltonian (\ref{HTJ}),
\begin{equation}
H_t = -\sum_{\langle ij \rangle_{\gamma}}\sum_{s\alpha\beta} 
t_{\gamma}^{\alpha\beta}\left(\hat{c}^{\dagger}_{is\alpha}\hat{c}_{js\beta}+
\text{H.c.}\right)-J_H\sum_i \bbox{S}_i^c \bbox{s}_i,
\label{HHP}
\end{equation}
which establishes the double-exchange mechanism in the correlated
system. Due to the strong Hund's coupling, core spins $\bbox{S}^c$ and
itinerant $e_g$ spins $\bbox{s}$ are not independent of each other;
rather a high-spin state with total onsite spin $S=S^c+\frac{1}{2}$
is formed. This unification of band and local spin subspaces suggests 
to decompose the $e_g$ electron into its spin and orbital/charge components. 
The $e_g$ spin can then be absorbed into the total spin, allowing an
independent treatment of spin and orbital/charge degrees of freedom
(see Fig.\ \ref{FIG:ABS}).
The procedure of this separation scheme is the following: In a first step
we introduce Schwinger bosons $d_{i\uparrow}$ and $d_{i\downarrow}$ 
(see, e.g., Ref.\ \onlinecite{AUE94}) to describe the $e_g$ spin
\begin{eqnarray*}
& s_i^+ = d^{\dagger}_{i\uparrow} d_{i\downarrow},\quad
s_i^- = d^{\dagger}_{i\downarrow} d_{i\uparrow},&\\
& s_i^z = \case{1}{2}(d^{\dagger}_{i\uparrow}d_{i\uparrow}-
d^{\dagger}_{i\downarrow}d_{i\downarrow}),&
\end{eqnarray*}
as well as Schwinger bosons $D^{\dagger}_{i\uparrow}$ and
$D^{\dagger}_{i\downarrow}$ to model the total onsite spin
\begin{eqnarray*}
& S_i^+ = D^{\dagger}_{i\uparrow} D_{i\downarrow},\quad
S_i^- = D^{\dagger}_{i\downarrow} D_{i\uparrow},&\\
& S_i^z = \case{1}{2}(D^{\dagger}_{i\uparrow}D_{i\uparrow}-
D^{\dagger}_{i\downarrow}D_{i\downarrow}).&
\end{eqnarray*}
These auxiliary particles are subject to the following constraints
that depend on the $e_g$ occupation number $n_i$:
\begin{eqnarray}
d^{\dagger}_{i\uparrow}d_{i\uparrow}+d^{\dagger}_{i\downarrow}d_{i\downarrow} 
&=& n_i,
\label{CON1}\\*
D^{\dagger}_{i\uparrow}D_{i\uparrow}+D^{\dagger}_{i\downarrow}D_{i\downarrow} 
&=& 2S-1+n_i.
\label{CON2}
\end{eqnarray}
The creation and destruction operators for $e_g$ electrons can then 
be expressed in terms of spinless fermions $c_{i\alpha}$ which carry
charge and orbital pseudospin and Schwinger bosons which carry spin: 
\[
c_{is\alpha} = c_{i\alpha} d_{is}.
\]
The kinetic-energy Hamiltonian (\ref{HHP}) now describes the 
transfer of pairs of spinless fermions and Schwinger bosons:
\begin{equation}
H_t = -\sum_{\langle ij \rangle_{\gamma}}\sum_{s\alpha\beta} 
t_{\gamma}^{\alpha\beta}\left(\hat{c}^{\dagger}_{i\alpha}\hat{c}_{j\beta}
d^{\dagger}_{is} d_{js}+\text{H.c.}\right)
-J_H\sum_i \bbox{S}_i^c \bbox{s}_i.
\label{HPP2}
\end{equation}
The Bose operators are subject to the constraint (\ref{CON1})
that enforces the operators
$d_{is}$ and $d^{\dagger}_{is}$ to act only on projected Hilbert 
spaces with one or zero Schwinger bosons, respectively.
Our aim is to absorb the $e_g$ spin into the total spin, which requires 
to map the $e_g$ operators $d_{is}$ onto operators $D_{is}$ for the
total spin. This is done by 
comparing the matrix elements of the two types of operators. 
On the one hand, keeping in mind that Hund's 
rule enforces the onsite spins to be always in a total-spin-symmetric 
state, the only nonvanishing matrix elements of the $d_{is}$  operators are
\begin{eqnarray}
\Big\langle S-\case{1}{2},m-\case{1}{2}\Big| 
d_{\uparrow}\Big|S,m\Big\rangle &=& \sqrt{(S+m)/(2S)},
\label{MA1}\\*
\Big\langle S-\case{1}{2},m+\case{1}{2}\Big| 
d_{\downarrow}\Big|S,m\Big\rangle &=& \sqrt{(S-m)/(2S)}.
\label{MA2}
\end{eqnarray}
In deriving the above expressions
we have used the Clebsch-Gordan coefficients 
$\langle S^c, m^c; m^e| S,m\rangle$ to decompose the
total-spin state $| S,m\rangle$ into core-  and $e_g$-spin 
states $|S^c, m^c;m^e\rangle$ with $m^e=\uparrow/\downarrow$.
These coefficients are given by
\begin{eqnarray*}
\langle S-\case{1}{2},m-\mbox{$\frac{1}{2}$};\uparrow|S,m\rangle &=&
\left[\frac{S+m}{2S}\right]^{1/2},\\*[5pt]
\langle S-\case{1}{2},m+\case{1}{2};\downarrow|S,m\rangle &=&
\left[\frac{S-m}{2S}\right]^{1/2}.
\end{eqnarray*}
On the other hand, the matrix elements of the $D_{is}$ operators are
\begin{eqnarray}
\Big\langle S-\case{1}{2},m-\case{1}{2}\Big| 
D_{\uparrow}\Big|S,m\Big\rangle &=& \sqrt{(S+m)},
\label{MA3}\\*
\Big\langle S-\case{1}{2},m+\case{1}{2}\Big| 
D_{\downarrow}\Big|S,m\Big\rangle &=& \sqrt{(S-m)}.
\label{MA4}
\end{eqnarray}
All other matrix elements vanish due to the constraint of 
Eq.\ (\ref{CON2}). By comparing Eqs.\ (\ref{MA1})-(\ref{MA2}) with 
Eqs.\ (\ref{MA3})-(\ref{MA4}) we obtain the mapping
\[
d_{is} = \frac{1}{\sqrt{2S}} D_{is}.
\]
Hamiltonian (\ref{HPP2}) can hence be rewritten in 
terms of total-spin operators $D_{is}$:
\begin{equation}
H_t = -\frac{1}{2S}\sum_{\langle ij \rangle_{\gamma}}\sum_{s\alpha\beta} 
t_{\gamma}^{\alpha\beta}\left(\hat{c}^{\dagger}_{i\alpha}\hat{c}_{j\beta}
D^{\dagger}_{is} D_{js}+
\text{H.c.}\right).
\label{HTB}
\end{equation}
The Hund's coupling term of Eq.\ (\ref{HPP2}) has been dropped here
as its presence is implied by the spin construction employed above.
This completes the separation of spin from the charge/orbital 
quantum numbers of $e_g$ electrons.

At low temperatures the magnetic moment of ferromagnetic manganites 
studied here is almost fully saturated. It is therefore reasonable to
expand Eq.\ (\ref{HTB}) around a ferromagnetic groundstate.
Technically this is done by condensing the spin-up Schwinger bosons 
(assuming the ferromagnetic moment to point along this direction)
and by treating spin-wave excitations around this groundstate in
leading order of $1/S$. Introducing magnon operators $b_i$,
the following relations hold:
\begin{eqnarray*}
D_{i\uparrow}   &=& \sqrt{2S-b^{\dagger}_i b_i} \approx
\sqrt{2S}\left(1-\frac{1}{4S}b^{\dagger}_i b_i\right),\\*
D_{i\downarrow} &=& b_i.
\end{eqnarray*}
This spin representation fixes the number of Schwinger
bosons per site to $2S$. The essence of the $1/S$ expansion
is to consider the presence of a hole as a small perturbation which 
changes the spin projection $S^z$  but not the spin magnitude
$S$. Employing magnon operators, the kinetic-energy 
Hamiltonian (\ref{HTB}) hence becomes
\begin{eqnarray}
H_t &=& -\sum_{\langle ij \rangle_{\gamma}} \sum_{\alpha\beta} 
t_{\gamma}^{\alpha \beta}\hat{c}^{\dagger}_{i\alpha} \hat{c}_{j\beta}
\nonumber\\*
&&+\frac{1}{2S}\sum_{\langle ij \rangle_{\gamma}}\sum_{\alpha\beta} 
t_{\gamma}^{\alpha \beta}
\hat{c}^{\dagger}_{i\alpha} \hat{c}_{j\beta} \Big(\case{1}{2}
b^{\dagger}_i b_i + \case{1}{2} b^{\dagger}_j b_j
-b^{\dagger}_i b_j \Big) \nonumber\\*
&&+ \text{H.c.}
\label{HKI}
\end{eqnarray}
The first term of Eq.\ (\ref{HKI}) describes the motion 
of strongly correlated fermions in a ferromagnetic background.
The second term controls the dynamics of spin excitations in the 
magnetic background and the interaction of these excitations 
with the fermionic sector.

At small magnon numbers, i.e., at low temperatures $T\ll T_C$,
Eq.\ (\ref{HKI}) can be mapped onto the following expression 
for the magnetic double-exchange bonds:
\begin{eqnarray}
H_t & = & -\sum_{\langle ij \rangle_{\gamma}} \sum_{\alpha\beta} 
t_{\gamma}^{\alpha \beta} \hat{c}^{\dagger}_{i\alpha} 
\hat{c}_{j\beta}\Big[\frac{3}{4} + \frac{1}{4S^2} \left(S_i^z S_j^z
+ S_i^- S_j^+\right)\Big]\nonumber\\*
&& + \text{H.c.}
\label{HKI2}
\end{eqnarray}
Equation (\ref{HKI2}) highlights an important point: The strength 
of double-exchange bonds is a fluctuating complex quantity. Only when 
treating the orbital and charge sectors on average, i.e., when
replacing the bond operators $\hat{c}^{\dagger}_{i\alpha} 
\hat{c}_{i\beta}$
by their mean-field value $\langle \hat{c}^{\dagger}_{i\alpha}
\hat{c}_{i\beta}\rangle$, an effective Heisenberg model as in a 
conventional mean-treatment of double exchange is obtained:
$H = J_{\text{DE}} \sum_{\langle ij\rangle} \bbox{S}_i
\bbox{S}_j$ with $J_{\text{DE}} = (2S^2)^{-1} \sum_{\alpha\beta} 
t_{\gamma}^{\alpha\beta}
\langle \hat{c}^{\dagger}_{i\alpha} \hat{c}_{j\beta}\rangle$. 
In Section \ref{SEC:MAG} we investigate in more detail the modification 
of the mean-field picture by fluctuations in the bond amplitude.

It is interesting to turn to the limit of classical 
spins shortly. Replacing the spin operators in Eq.\ (\ref{HKI2})
by their classical counterparts 
$S^z = S \cos\theta$ and $S^{\pm} = S \sin\theta e^{\mp i\phi}$, 
an effective fermionic model is obtained:
\begin{equation}
H_t = -\sum_{\langle ij \rangle_{\gamma}} \sum_{\alpha\beta} 
\tilde{t}_{\gamma}^{\alpha \beta} \hat{c}^{\dagger}_{i\alpha} 
\hat{c}_{j\beta}+ \text{H.c.}
\end{equation}
This model exhibits an unconventional phase-dependent
hopping amplitude:\cite{FTN1}
\[
\tilde{t}_{\gamma}^{\alpha \beta} = t_{\gamma}^{\alpha \beta}
\Big[\frac{3}{4} + \frac{1}{4} \left(\sin\theta_i 
\sin\theta_j+\sin\theta_i\sin\theta_j e^{i(\phi_i-\phi_j)} 
\right)\Big] .
\]
A similar result has been discussed in Refs.\ \onlinecite{MIL95,MUE96}
in terms of a Berry-phase effect.


\subsection{Superexchange Bonds}

At low- and intermediate-doping levels, virtual charge-transfer 
processes across the 
Hubbard gap becomes of importance. These superexchange processes 
establish an intersite interaction, which in the limit of a strong
Hund's coupling is described by [see Eq.\ (\ref{HTJ})]:
\begin{equation}
H_J = - J_{\text{SE}} \sum_{\langle ij \rangle_{\gamma}}
\left(\case{1}{4}-\tau_i^{\gamma}\tau_j^{\gamma}\right)
\left[\bbox{S}_i\bbox{S}_j + S(S+1)\right]n_i n_j.
\label{HJS}
\end{equation}
As mentioned above, superexchange is of ferromagnetic nature in 
the orbitally degenerate system with strong onsite correlations.
Double exchange and superexchange therefore act together 
in establishing the ferromagnetic exchange links in metallic 
manganites.\cite{FTN2}

Following the discussion on double-exchange bonds we express the spin 
operators in Eq.\ (\ref{HJS}) in terms of magnon operators $b_i$. This
leads to
\begin{eqnarray}
H_J &=& S J_{\text{SE}} \sum_{\langle ij \rangle_{\gamma}}
\left(\case{1}{4}-\tau_i^{\gamma}\tau_j^{\gamma}\right)
n_i n_j\nonumber\\*
&&\times\Big[\Big(\case{1}{2}b^{\dagger}_i b_i + \case{1}{2} 
b^{\dagger}_j b_j - b^{\dagger}_i b_j+\text{H.c.}\Big)- 
(2S+1)\Big].\nonumber\\*
\label{HJB}
\end{eqnarray}
Equation (\ref{HJB}) describes the interaction between
orbital fluctuations and the magnetic sector of the 
Hilbert space.\cite{FTN3} The phase dependence exhibited by 
the double-exchange counterpart Eq.\ (\ref{HKI}) is absent 
here. This is due to the fact that superexchange is a second-order 
process which depends only the amplitude but not on the phase of the
transfer amplitude.


\section{Magnon Dispersion}
\label{SEC:MAG}

In the previous section, the role of double-exchange and superexchange
processes in promoting ferromagnetic exchange bonds in manganites
was discussed. At intermediate-doping levels these exchange interactions
induce a ferromagnetic groundstate in a variety of manganese oxides. 
We now turn to analyze the propagation of magnetic excitations in this
ferromagnetic phase, namely by deducing the dispersion relation
of single-magnon excitations.

In a first step, we derive the correct operator for creating a magnetic 
excitation in hole-doped double-exchange systems.
It has to account for the fact that the total onsite spin
depends on whether a hole or an $e_g$ electron is present at that site: 
The spin number is $S-\frac{1}{2}$ in the former and $S$ in the latter 
case. This difference in the spin number was neglected
in the $1/S$ expansion employed in Sec.\ \ref{SEC:EXB}. Here this
approximation is no longer valid, which requires a rescaling of
the magnon operators $b_i$.
In general, a spin excitation is created by the operator $S_i^+$.
Expressing this operator in terms of Schwinger bosons
$S^+_i = D^{\dagger}_{i\uparrow}D_{i\downarrow}$, condensing
$D_{i\uparrow}$, and mapping $D_{i\downarrow}$ onto the magnon 
operator $b_i$, the following representation is obtained:
\[
S^+_i = \left\{
\begin{array}{ll}
\displaystyle
\sqrt{2S} \; b_i, & \text{for sites with $e_g$ electron},\\
\displaystyle
\sqrt{2S-1} \; b_i, & \text{for sites with hole}.
\end{array}\right.
\]
Assuming $S$ to be the ``natural'' spin number of the
system, the magnon operator $b_i$ hence has to be rescaled by a 
factor $[(2S-1)/(2S)]^{1/2}$ when being applied to hole sites:
\[
B_i = \left\{
\begin{array}{ll}
\displaystyle
b_i, & \text{for sites with $e_g$ electron},\\
\displaystyle
\sqrt{(2S-1)/(2S)} \; b_i, & \text{for sites with hole}.
\end{array}\right.
\]
The general magnon operator that automatically probes
the presence of an $e_g$ electron can finally be written as
\begin{eqnarray*}
B_i & = & b_i \Bigg[n_i + \sqrt{\frac{2S-1}{2S}} (1-n_i)\Bigg]
\nonumber\\*
& \approx & b_i-\frac{1}{4S} (1-n_i)\;b_i,
\end{eqnarray*}
where $n_i$ is the number operator of $e_g$ electrons.
$B_i$ represents the true Goldstone operator of hole-doped 
double-exchange systems. Its composite character comprises local and 
itinerant spin features which is a consequence of the fact that
static core and mobile $e_g$  electrons together form the total 
onsite spin. While the itinerant part of $B_i$ is of order $1/S$ only, 
it nevertheless is of crucial importance
to ensure consistency of the spin dynamics with the Goldstone
theorem, i.e., to yield an excitation mode whose energy vanishes
at zero-momentum.

Having derived the correct magnon operator for doped
double-exchange systems, we now study the propagation of 
the magnetic excitations it creates. The link between
sites that allows a local excitation to spread throughout 
the system is established by the exchange-bond Hamiltonians
(\ref{HKI}) and (\ref{HJB}). At low temperatures the
dynamics of spin waves which hence develop is captured by the 
single-magnon dispersion. The important question we are interested 
in is the following: To which extent is the magnon spectrum affected 
by fluctuations in the exchange bonds? To answer this question we 
express the full magnon spectrum $\tilde{\omega}_{\bbox{p}}$ in 
terms of the conventional mean-field dispersion $\omega_{\bbox{p}}$ 
and the magnon selfenergy $\Sigma(\omega,\bbox{p})$:
\begin{equation}
\tilde{\omega}_{\bbox{p}} =
\omega_{\bbox{p}}+\text{Re}[\Sigma(\omega_{\bbox{p}},\bbox{p})].
\label{WTI}
\end{equation}
Fluctuation are considered only on average in the former but are
explicitly accounted for in the latter term.
The mean-field dispersion $\omega_{\bbox{p}}$ as well as the 
scattering vertices needed to 
construct $\Sigma(\omega,\bbox{p})$ can be derived by commuting 
the magnon operator $B_i$ with the Hamiltonian. To be specific 
we explicitly perform this commutation, for now restricting ourselves to the 
double-exchange Hamiltonian $H_t$ given by Eq.\ (\ref{HKI}). In the
momentum representation we obtain
\begin{eqnarray}
[B_{\bbox{p}},H_t] & = & \omega_{\bbox{p}}B_{\bbox{p}} \nonumber\\*
&& + \frac{t}{2S} \sum_{\bbox{q}} \sum_{\alpha\beta}
A^{\alpha\beta}_{\bbox{p}}(\bbox{k})
\hat{c}^{\dagger}_{\bbox{k}\alpha} \hat{c}_{\bbox{k}-\bbox{q},\beta}
B_{\bbox{p}+\bbox{q}}.
\label{CM1}
\end{eqnarray}
The two terms on the r.h.s.\ of Eq.\ (\ref{CM1}) correspond to an 
expansion of the bond operators $\hat{c}^{\dagger}_{i\alpha} 
\hat{c}_{j\beta}$ around their average value:
\[
\hat{c}^{\dagger}_{i\alpha} \hat{c}_{j\beta}
\rightarrow
\langle \hat{c}^{\dagger}_{i\alpha} \hat{c}_{j\beta} \rangle
+\delta\left(
\hat{c}^{\dagger}_{i\alpha} \hat{c}_{j\beta}
\right).
\]
The mean-field magnon dispersion $\omega_{\bbox{p}}$ 
in the first term of Eq.\ (\ref{CM1}) is of conventional
nearest-neighbor Heisenberg form 
\begin{equation}
\omega_{\bbox{p}} = zD(1-\gamma_{\bbox{p}}),
\label{WBR}
\end{equation}
with the form factor $\gamma_{\bbox{p}} = z^{-1}\sum_{\bbox{\delta}} 
\exp(i\bbox{p}\bbox{\delta})$, $z=6$, and the 
spin-wave stiffness constant is $D=SJ_{\text{DE}}$. On this
mean-field level
the strength of the exchange bonds depends on the orbital
and charge degrees of freedom only on average:
$J_{\text{DE}}=(2S^2)^{-1} \sum_{\alpha\beta} 
t_{\gamma}^{\alpha\beta} \langle \hat{c}^{\dagger}_{i\alpha} 
\hat{c}_{j\beta}\rangle$. The second term in Eq.\ (\ref{CM1})
is the scattering vertex needed to construct the magnon selfenergy 
$\Sigma(\omega,\bbox{p})$. It describes the interaction between
magnons and orbital/charge fluctuations. The vertex 
function is
\[
A^{\alpha\beta}_{\bbox{p}}(\bbox{k}) = \gamma^{\alpha\beta}_{\bbox{k}} - 
\gamma^{\alpha\beta}_{\bbox{k}+\bbox{p}},
\]
with the form factor
$\gamma_{\bbox{k}}^{\alpha\beta} = (zt)^{-1}\sum_{\bbox{\delta}}
t_{\delta}^{\alpha\beta} \exp(i\bbox{k}\bbox{\delta})$. The 
vertex function $A^{\alpha\beta}_{\bbox{p}}(\bbox{k})$ vanishes
in the limit $\bbox{p} \rightarrow 0$ in compliance with
the Goldstone theorem.

Before we can engage in evaluating the
magnon selfenergy associated with the scattering vertex in 
Eq.\ (\ref{CM1}), the problem of dealing with the 
correlated nature of fermionic operators 
$\hat{c}^{\dagger}_{i\alpha}=c^{\dagger}_{is}(1-n_i)$ 
has to be addressed. To handle the constraint that allows only
for one $e_g$ electron per site, we employ an orbital-liquid 
scheme:\cite{ISH97,KIL98} Orbital and
charge degrees of freedom of the $e_g$ electron are treated on separate
footings by introducing ``orbiton'' and ``holon'' quasiparticles.
To describe an orbitally disordered state in which orbitals fluctuate
strongly, orbitons $f_i$ are assigned fermionic and holons $h_i$ bosonic 
statistics.\cite{KIL98} The original fermion operators are hence 
reexpressed by
\begin{equation}
c^{\dagger}_{i\alpha} = f^{\dagger}_{i\alpha} h_i.
\label{SOC}
\end{equation}
The local no-double-occupancy constraint is now relaxed to a global one:
\[
n^f_i+n^h_i = 1 \quad \rightarrow \quad \langle n^f_i \rangle + 
\langle n^h_i \rangle = 1.
\]
The main feature associated with the constrained nature of electrons,
namely the separation of energy scales of orbital and charge dynamics, 
sustains this procedure due to the fact that two different types of  
quasiparticles are being used. Introducing mean-field parameters
\begin{equation}
\chi=t^{-1}\sum_{\alpha\beta}t^{\alpha\beta}_{\gamma}
\langle f^{\dagger}_{i\alpha} f_{j\beta}\rangle,
\quad x = \langle b^{\dagger}_i b_j \rangle,
\label{MFP}
\end{equation}
where $x$ is the concentration of holes in the system, orbitons 
and holons can now be decoupled. We note that the two 
mean-field parameters in Eq.\ (\ref{MFP}) are approximately related by 
$\chi = \frac{1}{2}(1-x)$.

Employing representation (\ref{SOC}), we reexpress the 
commutator of Eq.\ (\ref{CM1}) in terms of orbiton and holon 
operators:
\begin{eqnarray}
[B_{\bbox{p}},H_t] &=&
\omega_{\bbox{p}}B_{\bbox{p}}\nonumber\\*
&&+ \frac{t}{2S} \sum_{\bbox{q}} \sum_{\alpha\beta}
C^{\alpha\beta}_{\bbox{p}}(\bbox{k}) 
f^{\dagger}_{\bbox{k}\alpha} f_{\bbox{k}-\bbox{q},\beta}
B_{\bbox{p}+\bbox{q}}\nonumber\\*
&&+\frac{t\chi}{2S} \sum_{\bbox{q}}
D_{\bbox{p}}(\bbox{k})
h_{\bbox{k}} h^{\dagger}_{\bbox{k}-\bbox{q}}
B_{\bbox{p}+\bbox{q}}.
\label{STJ}
\end{eqnarray}
The vertex functions are given by
\begin{eqnarray*}
C^{\alpha\beta}_{\bbox{p}}(\bbox{k}) &=& xA^{\alpha\beta}_{\bbox{p}}(\bbox{k}),\\
D_{\bbox{p}}(\bbox{k}) &=& \gamma_{\bbox{k}} - \gamma_{\bbox{k}+\bbox{p}}.
\end{eqnarray*}
Orbitons and holons have been decoupled in Eq.\ (\ref{STJ}) by 
employing the mean-field parameters $x$ and $\chi$ of
Eq.\ (\ref{MFP}). This yields two different types of scattering vertices,
one describing the interaction of magnons with orbital fluctuations, i.e., 
orbitons, the other of magnons with charge fluctuations, i.e., holons.

\begin{figure}
\begin{minipage}[b]{0.32\linewidth}
\centering
\epsfig{file=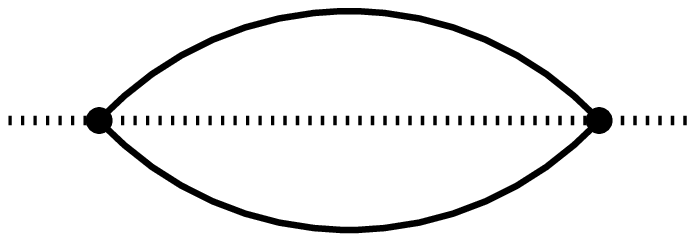,width=\linewidth}\\[3pt]
(a)
\end{minipage}
\hfill
\begin{minipage}[b]{0.32\linewidth}
\centering
\epsfig{file=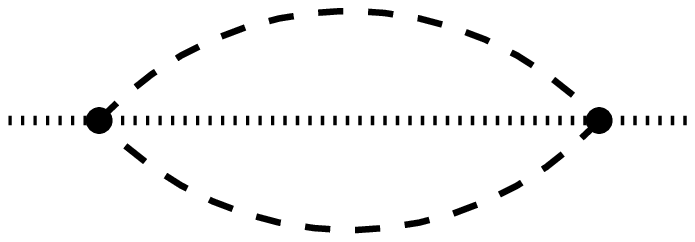,width=\linewidth}\\[3pt]
(b)
\end{minipage}
\hfill
\begin{minipage}[b]{0.32\linewidth}
\centering
\epsfig{file=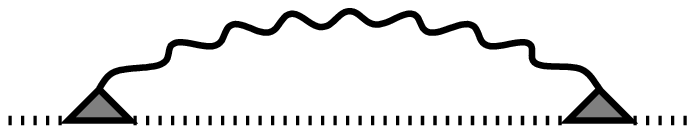,width=\linewidth}\\[3pt]
(c)
\end{minipage}\\[6pt]
\caption{Magnon selfenergies describing the effect of magnon scattering on 
(a) orbital fluctuations, (b) charge fluctuations, and (c) phonons. Solid, 
dashed, dotted, and wiggled lines denote orbiton, holon, magnon, and
phonon propagators, respectively.}
\label{FIG:SCO}
\end{figure}

Finally we include in our treatment the magnetic bonds stemming
from superexchange processes as described by $H_J$ in Eq.\ (\ref{HJB}).
The effect is twofold: Superexchange enhances the spin-wave stiffness 
$D$ which now becomes
\[
D = S(J_{\text{DE}}+J_{\text{SE}}) = t\chi(x+x_0)/(2S),
\]
with $x_0 = 2\chi t/U_1$; further, superexchange processes renormalize 
the vertex function of magnon-orbiton scattering which becomes
\[
C^{\alpha\beta}_{\bbox{p}}(\bbox{k})=
x A^{\alpha\beta}_{\bbox{p}}(\bbox{k})+x_0 B^{\alpha\beta}_{\bbox{p}}
(\bbox{k},\bbox{q}),
\]
with
\[
B^{\alpha\beta}_{\bbox{p}}(\bbox{k},\bbox{q}) =
\gamma^{\alpha\beta}_{\bbox{k}} + \gamma^{\alpha\beta}_{\bbox{k}-\bbox{q}} - 
\gamma^{\alpha\beta}_{\bbox{k}+\bbox{p}} -  
\gamma^{\alpha\beta}_{\bbox{k}-\bbox{q}-\bbox{p}}.
\]
From the two types of scattering vertices in Eq.\ (\ref{STJ}), two 
contributions to the magnon selfenergy follow. 
These describe the scattering of magnons on orbitons 
and on holons and are depicted in Figs.\ \ref{FIG:SCO}(a) and (b),
respectively.

An important piece of physics is still missed in the above treatment,
namely the Jahn-Teller coupling of orbitals to the lattice.\cite{MIL96}
In a cubic system
there exist two independent Jahn-Teller modes $Q_2$ and $Q_3$ which
lift the degeneracy of singly occupied $e_g$ orbitals. The interaction 
between orbitals and these two orthogonal lattice modes 
is described by
\begin{equation}
H_{\text{JT}} = -\sum_i \left(g_2Q_{2i} \sigma_i^x + g_3Q_{3i}
\sigma^z_i\right),
\label{HOL}
\end{equation}
where the Pauli matrices $\sigma^{x/z}_i$ act on the orbital 
subspace and the coupling constants $g_2\approx g_3$. The crystal
dynamics is controlled by the Hamiltonian
\begin{equation}
H_{\text{ph}} = \frac{K}{2} \sum_i \bbox{Q}_i^2 +
K_1 \sum_{\langle ij\rangle_{\gamma}} Q_i^{\gamma} Q_j^{\gamma}
+ \frac{1}{2M}\sum_i \bbox{P}^2_i,
\label{PDI}
\end{equation}
with $Q_i^{x/y} = (Q_{3i}\pm \sqrt{3} Q_{2i})/2$, $Q_i^z = Q_{3i}$, and
$\bbox{Q}_i = (Q_{2i},Q_{3i})$; $\bbox{P}_i$ denotes the conjugate
momentum vector corresponding to the lattice distortions $\bbox{Q}_i$.  
The three terms on the r.h.s.\ of Eq.\ (\ref{PDI}) account for the 
crystal deformation energy, the correlations between neighboring sites, 
and the lattice kinetics, respectively. Equation (\ref{PDI}) can be 
diagonalized in the momentum representation, yielding
\begin{equation}
H_{\text{ph}} = \sum_{\bbox{k}\nu} \omega^{\nu}_{\bbox{k}}
a^{\dagger}_{\bbox{k}\nu} a_{\bbox{k}\nu},
\end{equation}
with index $\nu=\pm$ and the phonon dispersions
\begin{equation}
\omega^{\pm}_{\bbox{k}} = \omega_0
\left(\kappa_{1\bbox{k}}\pm\sqrt{\kappa_{2\bbox{k}}^2+\kappa_{3\bbox{k}}^2}
\right)^{1/2}.
\label{PHD}
\end{equation}
Here, $\kappa_{1\bbox{k}} = 1+k_1(c_x+c_y+c_z)$, 
$\kappa_{2\bbox{k}} = k_1 \eta^{(2)}_{\bbox{k}}$,
$\kappa_{3\bbox{k}} = k_1\eta^{(3)}_{\bbox{k}}$ with $k_1 = K_1/K$ and
$\eta^{(2)}_{\bbox{k}} = -\sqrt{3}(c_x-c_y)/2$,
$\eta^{(3)}_{\bbox{k}} = c_z-\frac{1}{2}c_x-\frac{1}{2}c_y$ with
$c_{\alpha}= \cos k_{\alpha}$, and $\omega_0=\sqrt{K/M}$.

While there is no direct coupling between spins and phonons in the present 
system, lattice modes nevertheless strongly affects the spin dynamics. 
The link between spin and lattice is established via the orbital channel:
The coupling of orbitals to the lattice imposes low phononic frequencies 
onto orbital fluctuations. This acts to enhance the modulation of magnetic 
exchange bonds; thereby the effect of phonons extends onto the spin 
sector. To study this mechanism in more detail, we construct an effective 
spin-phonon-coupling Hamiltonian from which we then calculate the 
phononic contribution to the magnon selfenergy. Combining the 
spin-orbital-coupling term of the exchange Hamiltonians (\ref{HKI}) 
and (\ref{HJB}) with the orbital-lattice Hamiltonian (\ref{HOL}) we 
obtain (see Fig.\ \ref{FIG:MPS}):
\begin{figure}
\centering
\epsfig{file=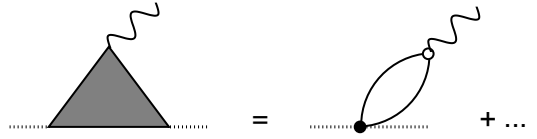,width=0.8\linewidth}\\[6pt]
\caption{Effective spin-phonon-coupling vertex. The dominant 
contribution shown on the right stems from a combination of 
spin-orbital- (filled dot $\propto t$)
and orbital-lattice- (open dot $\propto g_2$) coupling vertices.
The orbital susceptibility depicted by a bubble controls the 
coupling strength. Solid, dotted, and wiggled lines represent orbiton, 
magnon, and phonon propagators, respectively.}
\label{FIG:MPS}
\end{figure}
\begin{equation}
H_{\text{s-ph}} = -\sum_{\bbox{p}\bbox{q}\nu}
g_{\bbox{p}\bbox{q}}^{\nu}
(a^{\dagger}_{\bbox{q}\nu} + a_{\bbox{q}\nu}) 
B^{\dagger}_{\bbox{p}} B_{\bbox{p}+\bbox{q}}.
\label{HSL}
\end{equation}
The coupling constants in Eq.\ (\ref{HSL}) are
\begin{eqnarray*}
g_{\bbox{p}\bbox{q}}^+ &=&
\epsilon_0\left(\frac{\omega_0}{\omega_{\bbox{q}}^+}\right)^{1/2}
\left(\lambda_{\bbox{p}\bbox{q}}^{(3)}\cos\Theta_{\bbox{q}} -
\lambda_{\bbox{p}\bbox{q}}^{(2)}\sin\Theta_{\bbox{q}}\right),\\*
g_{\bbox{p}\bbox{q}}^- &=& 
\epsilon_0\left(\frac{\omega_0}{\omega_{\bbox{q}}^-}\right)^{1/2}
\left(\lambda_{\bbox{p}\bbox{q}}^{(3)}\sin\Theta_{\bbox{q}} +
\lambda_{\bbox{p}\bbox{q}}^{(2)}\cos\Theta_{\bbox{q}}\right),
\end{eqnarray*}
with $\epsilon_0 = (E_{\text{JT}}a_0^2\omega_0/S^2)^{1/2}$ and
$\lambda^{(\alpha)}_{\bbox{p}\bbox{q}} = 
(\eta^{(\alpha)}_{\bbox{q}} - \eta^{(\alpha)}_{\bbox{p}}-
\eta^{(\alpha)}_{\bbox{p}+\bbox{q}})$. Further
\begin{eqnarray*}
\cos\Theta_{\bbox{q}} &=& \frac{1}{\sqrt{2}}
\left(1+\frac{\kappa_{3\bbox{q}}}{\sqrt{\kappa_{2\bbox{q}}^2+
\kappa_{3\bbox{q}}^2}}\right)^{1/2},\\*
\sin\Theta_{\bbox{q}} &=& \frac{1}{\sqrt{2}}
\left(1-\frac{\kappa_{3\bbox{q}}}{\sqrt{\kappa_{2\bbox{q}}^2+
\kappa_{3\bbox{q}}^2}}\right)^{1/2}\text{sign}(\kappa_{2\bbox{q}}).
\end{eqnarray*}
The strength of the spin-lattice interaction is controlled by the 
orbital susceptibility $\langle (f^{\dagger}_{i+z,\uparrow} f_{i\uparrow}) 
(\sigma_i^z)\rangle_{\omega}$ which enters 
the parameter $a_0 = t(x+x_0)\langle (f^{\dagger}_{i+z,\uparrow} 
f_{i\uparrow}) (\sigma_i^z)\rangle_{\omega=0}$; the zero-frequency
limit is admissible bearing in mind that the energy scale of orbital 
fluctuations exceeds the one of phonons. The phononic contribution 
to the magnon selfenergy that follows from Hamiltonian (\ref{HSL}) 
can finally be calculated, the corresponding diagram is depicted 
in Fig.\ \ref{FIG:SCO}(c).


\section{Comparison with Experiment}

We are now in the position to evaluate the selfenergies of 
Fig.\ \ref{FIG:SCO}. 
Charge and orbital susceptibilities are calculated using mean-field Green's 
functions in slave-boson $h_i$ and fermion $f_i$ subspaces. For the spectral 
density of Jahn-Teller phonons in Fig.\ \ref{FIG:SCO}(c) we employ
the expression
\begin{equation}
\rho^{\text{ph}}_{\pm}(\omega,\bbox{q}) = 
\frac{1}{\pi}\frac{\omega}{\omega_{\bbox{q}}^{\pm}}
\frac{\Gamma}{(\omega-\omega_{\bbox{q}}^{\pm})^2+\Gamma^2},
\end{equation}
which phenomenologically accounts for the damping $\Gamma$ of phonons
due to their coupling to orbital fluctuations. The phonon dispersion
$\omega_{\bbox{q}}^{\pm}$ is given by Eq.\ (\ref{PHD}).

\begin{figure}
\centering
\epsfig{file=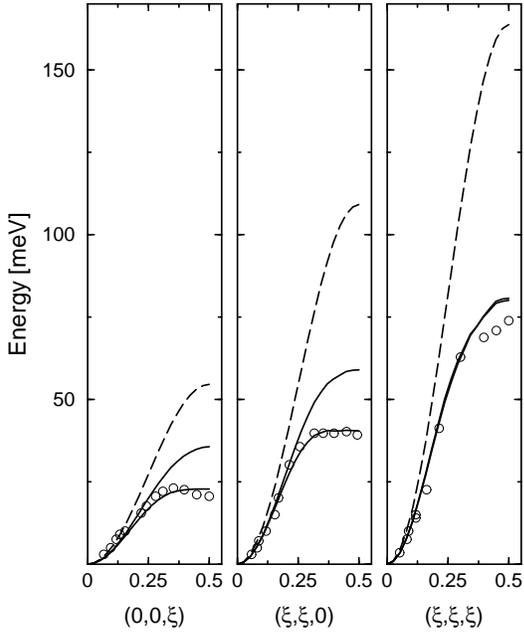,width=0.8\linewidth}\\[6pt]
\caption{Magnon dispersion along $(0,0,\xi)$, $(\xi,\xi,0)$, and
$(\xi,\xi,\xi)$ directions, where $\xi=0.5$ at the cubic zone boundary. 
Experimental data from Ref.\ \protect\onlinecite{HWA98} are 
indicated by circles, the 
mean-field dispersion $\omega_{\bbox{p}}$ of Eq.\ (\protect\ref{WBR})
is marked by a dashed line; the latter is of conventional 
nearest-neighbor Heisenberg form. Solid lines represent the theoretical 
result for the dispersion $\tilde{\omega}_{\bbox{p}}$ defined by 
Eq.\ (\protect\ref{WTI});
it includes charge, orbital, and lattice effects. The upper curve is 
obtained for dispersionless phonons with $k_1=0$, the lower one is a
fit to the experimental data with $k_1=-0.33$ corresponding to
ferrotype orbital-lattice correlations.}
\label{FIG:MRN}
\end{figure}

The expressions obtained from the diagrams in Fig.\ \ref{FIG:SCO} contain 
summations over momentum space which we perform numerically using a 
Monte-Carlo algorithm. The result is shown by solid lines in 
Fig.\ \ref{FIG:MRN}. 
For comparison, the experimental data of Ref.\ \onlinecite{HWA98}
are marked by circles and the bare mean-field dispersion 
$\omega_{\bbox{p}}$ is indicated by a dashed line. 
The following parameters are chosen: The hopping amplitude $t=0.4$~eV 
is adjusted to fit the spin stiffness in 
Pr$_{0.63}$Sr$_{0.37}$MnO$_3$;\cite{HWA98}
further we use $U_1=4$~eV.\cite{FEI99}
The phonon contribution depends on the quantities $E_{\text{JT}} a_0^2 
\equiv (g_2a_0)^2/2K = 0.004$~eV,\cite{FTN4}
$\omega_0 = 0.08$~eV,\cite{FTN5} and $\Gamma=0.04$~eV.

The upper solid line in Fig.\ \ref{FIG:MRN} is obtained for
$k_1 = 0$. In this case intersite orbital-lattice correlations
in Hamiltonian (\ref{PDI}) are discarded~--- 
phonons are dispersionless. A pronounced 
softening of magnons at large momenta can be observed. 
A more detailed analysis reveals this effect to be mostly
due to fluctuations of the orbital and lattice degrees of 
freedom. In contrast, charge fluctuations are found to play 
only a minor role. We attribute this to the fact that the spectral 
density of charge fluctuations lies well above the magnon band. 
Orbital and lattice fluctuations, on the other hand, are of
rather low frequency ($\propto xt$ and $\propto \omega_0^{\text{ph}}$,
respectively) and hence affect the spin-wave dispersion in a more
pronounced way.

\begin{figure}
\centering
\epsfig{file=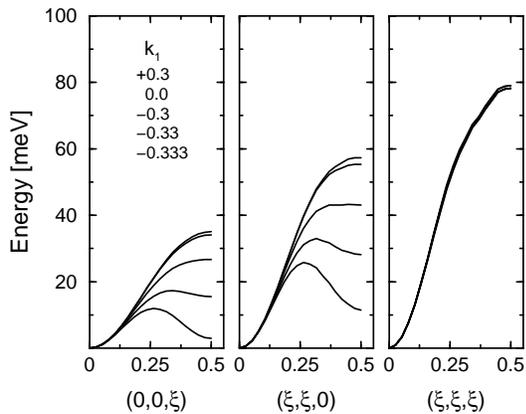,width=0.8\linewidth}\\[6pt]
\caption{Magnon dispersion including charge, orbital, and
lattice effects. Different values for $k_1$ controlling
intersite orbital-lattice correlations are used. The softening 
enhances as $k_1\rightarrow -\frac{1}{3}$ corresponding to an
instability point towards ferrotype orbital-lattice order.}
\label{FIG:PLK}
\end{figure}

The lower solid line in Fig.\ \ref{FIG:MRN} is obtained for
$k_1=-0.33$ which yields a fit to the experimental 
data of Ref.\ \onlinecite{HWA98}. The directional dependence of the
magnon renormalization seen in experiment is well 
reproduced: The effect is strongest in 
$(0,0,\xi)$ and $(0,\xi,\xi)$ directions. A key observation here
is the crucial role of intersite correlations of orbital-lattice 
distortions~--- these are captured by the phononic dispersion 
being controlled by the parameter $k_1$. In order to reproduce the
experimental data we are forced to assume these correlations to be 
of {\it ferrotype}, i.e., $k_1<0$. We believe this somewhat
surprising result to reflect an important piece of new physics:
Conventionally one would expect $k_1>0$ 
associated with a tendency of the orbital/lattice sector to develop 
antiferrotype order.\cite{KUG82} In the hole-doped system, however, 
this effect competes against charge mobility which prefers a ferrotype 
orbital orientation. The latter allows to minimize the kinetic energy by 
maximizing the transfer amplitude between sites. While Jahn-Teller 
lattice effects prevail at low doping, we speculate the kinetic 
energy to dominate at large enough hole concentrations. 
In fact, low-dimensional ferrotype orbital correlations 
(resonating $|x^2-y^2\rangle$, $|x^2-z^2\rangle$, $|y^2-z^2\rangle$
planar configurations) have been observed to evolve in a bosonic 
description of orbital fluctuations.\cite{ISH97} The fermionic 
description of orbitals employed in the present work emphasizes on 
modeling a strongly fluctuating orbital-liquid state, but underestimates 
these orbital-lattice instabilities. In order to simulate the competition 
between Jahn-Teller 
effect and kinetic energy we therefore turn to a phenomenological approach:
By tuning the parameter $k_1$ we control the character of intersite 
orbital-lattice correlations. The result for different values
of $k_1$ is shown in Fig.\ \ref{FIG:PLK} where 
$E_{\text{JT}}a_0^2=0.006$~eV is used.

Ferrotype orbital correlations with $k_1<0$ are found to
be most effective in renormalizing the magnon spectrum. 
This is ascribed to slowly fluctuating layered orbital configurations 
which effectively reduce the dimensionality of exchange bonds.
We note that magnons in $(\xi,\xi,\xi)$ direction are sensible to 
all three spatial directions of the exchange bonds; their dispersion
therefore remains unaffected by the local symmetry breaking 
induced by low-dimensional orbital correlations.
As an instability towards orbital-lattice order is 
approached, exchange-bond fluctuations become quasistatic. 
In the magnon spectrum this is reflected by a strong enhancement 
of the renormalization effect as is seen in Fig.\ \ref{FIG:PLK} for 
$k_1\rightarrow -\frac{1}{3}$. The layered orbital structure which 
evolves at this point is accompanied by a layered spin 
structure; the latter is indeed experimentally observed at doping 
levels of about $x=0.5$.\cite{KAW97,MOT98}

We finally note that the softening of magnons at the zone boundary 
leads to a reduction of $T_C$. Remarkably, the small-$q$ spin 
stiffness $D$ remains unaffected which explains the 
anomalous enhancement of the $D/T_C$ ratio in low-$T_C$ 
manganites.\cite{FER98}


\section{Conclusion}

In summary, we have presented a theory of the spin dynamics in
ferromagnetic manganites. Taking into account the orbital 
degeneracy and the correlated nature of $e_g$ electrons,
we analyzed the structure of magnetic exchange bonds; these 
are established by the intersite transfer of electrons 
in coherent double-exchange and virtual superexchange
processes. Orbital and charge fluctuations are shown
to strongly modulate the exchange bonds, leading to a 
softening of the magnon excitation spectrum close to the
Brillouin zone boundary. The presence of Jahn-Teller 
phonons further enhances the effect. This peculiar interplay
between double-exchange physics and orbital-lattice
dynamics becomes dominant close to the instability towards 
an orbital-lattice ordered state. The unusual 
magnon dispersion experimentally observed in low-$T_C$
manganites can hence be understood as a precursor effect of
orbital-lattice ordering. While the
softening of magnons at the zone boundary is responsible
for reducing the value of $T_C$, the small-momentum spin 
dynamics that enters the spin-wave stiffness $D$ remains 
virtually unaffected. This explains the enhancement of
the ratio $D/T_C$ observed in low-$T_C$ compounds. In general
it can be concluded that strong correlations and orbital fluctuations
play a crucial role in explaining the peculiar magnetic properties of 
metallic manganites.


We would like to thank H.\ Y.\ Hwang and P.\ Horsch for 
stimulating discussions.



\begin{references}
\bibitem{FTN1} It is noticed that the phase-dependent part of the
fermionic hopping is only of order $1/S$ [see Eq.\ (\protect\ref{HKI})].
\bibitem{FTN2} Here we do not discuss the superexchange between
core spins as well as finite-$J_H$ effects which are
responsible for a variety of antiferromagnetic structures
in manganites. These effects are only of minor importance
in the cubic metallic phase under consideration.
\bibitem{FTN4} A mean-field calculation gives $a_0\approx 0.1$. Our 
fitting then implies a reasonable Jahn-Teller binding energy 
$E_{\text{JT}} \approx 0.4$~eV.
\end{references}
\end{document}